\documentclass[twocolumn,twoside,prd,floatfix,letterpaper]{revtex4}

\usepackage{ifpdf}
\ifpdf 
  \usepackage[pdftex]{graphicx} 
\else 
  \usepackage{graphicx} 
\fi
\usepackage{amsmath}
\usepackage{amssymb}
\usepackage{fancyhdr}
\usepackage{color}
\usepackage{rotating}
\usepackage{natbib}
\usepackage[colorlinks,hyperindex]{hyperref}
\usepackage{breakurl}

\begin{document} 

\title{The Solution to Science's Replication Crisis}
\author{Bruce Knuteson}
\homepage{http://bruceknuteson.com}
\email{knuteson@kn-x.com}

\begin{abstract}
The solution to science's replication crisis is a new ecosystem in which scientists sell what they learn from their research.  In each pairwise transaction, the information seller makes (loses) money if he turns out to be correct (incorrect).  Responsibility for the determination of correctness is delegated, with appropriate incentives, to the information purchaser.  Each transaction is brokered by a central exchange, which holds money from the anonymous information buyer and anonymous information seller in escrow, and which enforces a set of incentives facilitating the transfer of useful, bluntly honest information from the seller to the buyer.  This new ecosystem, {\em{capitalist science}}, directly addresses socialist science's replication crisis by explicitly rewarding accuracy and penalizing inaccuracy.  

The problematic incentives within socialist science that have led to socialist science's replication crisis are too embedded to remove.  Those seeking accurate knowledge to inform important, high stakes decisions will gradually start to rely on capitalist science, which penalizes the selling of information that turns out to be wrong.  Researchers with genuinely valuable information will choose to sell it rather than publish it away for free.  The most useful, interesting, and accurate information will therefore gradually move from the ecosystem of socialist science to the ecosystem of capitalist science, where it is properly valued.  

Capitalist science is the solution to socialist science's replication crisis.
\end{abstract}

\maketitle

\section{Socialist Science}
\label{sec:SocialistScience}

Drug company Q reads a scholarly article in a distinguished journal authored by academic A.  Motivated in significant part by A's findings, Q starts a new research effort in this seemingly promising direction.

One year and one million dollars later, Q realizes A's findings are basically wrong~\cite{begley2012drug,prinz2011believe,freedman2015economics,chalmers2009avoidable,scott2008design,begley2013unappreciated,steckler2015preclinical,de2015failure,kola2004can,macleod2014biomedical,kyzas2007almost,hirst2014need,Hyman155cm11,miller2010pharma,schoenfeld2013everything}.

Variants of this story may be invoked by a convenient buzz phrase:  science's ``replication crisis,'' ``replicability crisis,'' or ``reproducibility crisis.''  Q is not always a drug company.  A is usually an academic.

Q, the sucker in this story, realizes he is continually getting shafted~\cite{ioannidis2005most,ioannidis2013s,pankevich2014improving,lowenstein2009uncertainty,henderson2013threats,dirnagl2012international,dirnagl2009stroke,dirnagl2006bench,rosenblatt2016incentive}.  Systematic attempts to repeat many published studies fail~\cite{mobley2013survey,open2015estimating,ioannidis2005contradicted,ioannidis2007non,steward2012replication,chang2015economics}, spreading awareness to the masses and stoking discontent.  Debates erupt over the scale of the problem~\cite{maxwell2015psychology,anderson2016response,stroebe2014alleged,klein2014investigating,pashler2012replicability,camerer2016evaluating,etz2016bayesian,michalek2010costs}.  Some hope extermination of psychology departments will be sufficient~\cite{fanelli2010positive,lilienfeld2012public,cesario2014priming,Pashler01112012,bakker2012rules,Bones01052012,wagenmakers2011psychologists,ferguson2015everybody}.  Others work to re(de)fine the notion of ``replication''~\cite{goodman2016does,brandt2014replication,clemens2015meaning}.  Incremental improvements to the current system are proposed~\cite{ioannidis2014make,collins2014nih,vesterinen2010improving,asendorpf2013recommendations,nosek2015promoting,cumming2013new,iqbal2016reproducible,holdren2013increasing,miguel2014promoting,roche2014troubleshooting,chalmers2014increase,ioannidis2014increasing,salman2014increasing,yordanov2015avoidable,moher2015four,ioannidis2014assessing,vasilevsky2013reproducibility,international2004international,kilkenny2010improving,macleod2009reprint,tooth2005quality,festing2002guidelines,moher2010guidance,casadevall2012reforming,nosek2012scientificI,nosek2012scientificII,everett2015tragedy,nekrutenko2012next,sandve2013ten,ioannidis2011improving,pusztai2013reproducibility,valentine2011replication,kidwell2016badges,koole2012rewarding,landis2012call,moher2010consort,moseley2014beyond,peers2012search,peers2014can,ram2013git,schooler2014metascience}~\footnote{{\em{Open science}}, or {\em{even-more-socialist science}}, is a common theme among proposed incremental improvements.  The idea of encouraging more transparent access to data and analysis code is an attractive one.  We ourselves pushed it in particle physics -- hard, and for many years.  Unfortunately, if unsurprisingly, incentives are simply too misaligned for it to work.  If we, as a society, want more of something -- like apples, say, or knowledge about how nature works -- we may be better off making it easy for people who produce apples to sell them than mandating that they make their orchards available for anyone to pick.}.  Most are never enacted~\cite{baker2014two}.  The rest have limited impact~\cite{vanpaemel2015we,roche2015public,vines2014availability,wicherts2006poor,moher2016increasing,grant2013reporting,joseph2013open,song2010dissemination,prayle2012compliance,clarke2010clinical,florez2016bias,bramhall2015quality,plint2006does}.

The incentive structures contributing to the replication crisis are observed to be deeply entrenched features of the ecosystem within which science is done~\cite{begley2015reproducibility,simmons2011false,fiedler2011voodoo,ferguson2012vast,young2008current,tsilidis2013evaluation,brembs2013deep,steen2013has,oboyle2014chrysalis,stamatakis2013undue,sena2010publication,mathieu2009comparison,hannink2013comparison,crowe2015patients,blumle2016fate,chan2014increasing,glasziou2014reducing,glasziou2014role,hoffmann2012scatter,franco2014publication,button2013power,dwan2008systematic,ioannidis2008most,van2010can,ioannidis2012science,stroebe2012scientific,wagenmakers2012agenda,john2012measuring,fanelli2011negative,alberts2014rescuing,ioannidis2014publication,chambers2014instead} and subsequently applied~\cite{glasziou2005paths,duff2010adequacy,dancey2010quality,kilkenny2009survey,mcglynn2003quality,lemon2005surveying,ioannidis2001completeness,savovic2012influence,turner2008selective,bero1998closing,ramagopalan2014prevalence,williamson2012driving}.  This ecosystem comprises institutions (including funding agencies, industry, universities, and academic journals), culture, accepted practice (including procedures by which grant money is awarded, articles published, promotions granted, and accolades bestowed), and the bureaucracy required to support all of these.  The ecosystem is complex, multifaceted, and not easily changed.  The incentive structures contributing to the replication crisis are, similarly, not easily changed.  A timely solution to the replication crisis probably requires a new ecosystem.

For convenience, we refer to the current ecosystem~\cite{stephan1996economics} -- funded by taxpayers, with published results available to all citizens -- as {\em{socialist science}}~\footnote{The phrase ``socialist science'' is obviously a grotesquely crude caricature of the intricate and often nuanced set of incentives joining the actors in the current science ecosystem.  We intend the phrase as a neutral description of an aspect of the current ecosystem germane to the present discussion.  A reader who dislikes the phrase is encouraged to mentally replace it with ``the ecosystem within which science is currently carried out,'' or some alternative reference thereto.}~\footnote{We focus on the incentive flaws common to all socialist science, ignoring differences among scientific disciplines.  Rather than treat the symptoms, which express differently in the social sciences, life sciences, and physical sciences, we focus on the underlying disease.}.

\section{Capitalist Science}
\label{sec:CapitalistScience}

Perhaps we can construct a different ecosystem.  

Lacking imagination, we look to see what other professions do.  Many professions exhibit a very interesting behavior.  There occur {\em{pairwise transactions}}, in which one party {\em{sells}} something of value to another party.  One party (A) gives something of value to some other party (Q).  In return, Q gives A something called {\em{money}}.

Perhaps our academic (A) can sell what he learns to those (Q) who find A's information valuable.  Many deep pocketed industries, including the pharmaceutical industry, make high stakes decisions motivated by academic research.  A should have little difficulty finding customers Q.

To incentivize accuracy, we need some sort of transaction that causes A to lose money if he turns out to be wrong.  We therefore need transactions for which the (in)accuracy of A's information can eventually be objectively determined, and we need a reliable, low-cost procedure for making that determination.

Suppose Q has a specific question and is willing to pay for a useful answer.  Q can ensure the usefulness of any answer he receives by specifying, explicitly or categorically, the set of allowed possible answers.  A third party (X) brokers the transaction, holds money from Q and A in escrow, and acts as arbiter.  If A provides an answer outside Q's set of allowed possible answers, A loses money.

In many cases of practical interest, Q can be tasked with determining the accuracy of A's answer.  If Q wants to know what chemical matter binds to a particular target and A answers with a specific molecule, Q will eventually know whether A's answer was right or wrong, and Q can provide evidence to this effect to arbiter X.  X collects a deposit from Q at the beginning of the transaction, and X returns this deposit to Q when Q provides sufficient evidence by a previously agreed upon date.  Q pays the same amount whether A's answer turns out to be right or wrong.  X's fee for brokering the transaction is independent of whether A's answer turns out to be right or wrong and whether or not Q provides evidence X considers sufficient.  A has an explicit monetary incentive to be accurate and an explicit monetary disincentive to provide information unless he is pretty sure he is correct.  Q has an explicit monetary incentive to ask a question that will be directly useful to him, and for which he can eventually determine, with reasonable proof, the accuracy of whatever answer A provides.  X, as an ongoing business interest, has a clear monetary incentive to thoughtfully serve its role as a reasonable and objective adjudicator of the evidence Q provides.  Q, A, and X all have an interest in ensuring Q's question is carefully specified to avoid subsequent ambiguities as to A's correctness.  This transaction protocol enforces a set of incentives facilitating paid, arm's length transfer of useful, bluntly honest information from A to Q.

Using this transaction protocol, introduced for the first time in Ref.~\cite{knuteson2016knowledge} and implemented at Ref.~\footnote{Kn-X; \url{http://kn-x.com}.  Patent pending.}, scientists can sell what they learn from their research.  

For convenience, we refer to this new ecosystem as {\em{capitalist science}}~\footnote{The phrase ``capitalist science'' is intended as a neutral description of a salient feature of this new ecosystem.  A reader who dislikes the phrase is encouraged to mentally replace it with ``the new ecosystem,'' or something similar.}~\footnote{The ``capitalist science'' in this article supersedes that of Ref.~\cite{knuteson2011capitalist}, which in retrospect is more of a hybrid between socialist and capitalist science.}~\footnote{To be clear, our use of the term ``capitalist'' does not arise from a belief, seemingly held by many, that free markets are the optimal solution to all problems, nor from a belief that global financial markets at the time of this writing function well and should be emulated.  (You have no idea.)  The complexity and frequent opacity of today's capitalism highlights the glaring need for a mechanism facilitating useful, bluntly honest information transfer between remote parties.   Given the embarrassing, hidden-in-plain-sight, farcically tragic comedy of errors that is recent financial history, a mechanism facilitating useful, bluntly honest, arm's length information transfer may turn out to be our best shot at saving capitalism \ldots\ or at least at sending a few of those responsible to prison next time around.}.

In capitalist science, A makes money only if he turns out to be correct.  A loses money if he turns out to be wrong.  Every transaction includes a monetary incentive for Q to determine the accuracy of A's answer and to back this determination with evidence deemed sufficient by an objective third party (X).  These features directly address socialist science's reproducibility crisis.  These incentives, present in capitalist science, explicitly and directly reward accuracy and penalize inaccuracy.  The absence of such explicit and direct incentives in socialist science is the reason socialist science is suffering a reproducibility crisis.

Capitalist science is the solution to socialist science's reproducibility crisis.

\section{Summary}
\label{sec:Summary}

Now that the solution to socialist science's reproducibility crisis has been found, we can roll up our sleeves \ldots\ sit back, relax, and let events unfold.

Socialist science will continue much as it has, producing results of similar quality.  Serious, thoughtful, and impassioned attempts will be made to save it.  Some of these may appear to work for a while, but ultimately they will fail.  The reproducibility crisis has been stewing for decades~\cite{lykken1968statistical,elms1975crisis,greenwald1975consequences,rosenthal1979file,altman1994scandal,hackam2006translation,pocock1987statistical,sterling1995publication,vul2009puzzlingly,easterbrook1991publication,kerr1998harking}.  The relevant malincentives are too firmly embedded in its large and unwieldy ecosystem for socialist science to produce a meaningful solution from within.

The key to change is Q, the socialist science sucker from Section~\ref{sec:SocialistScience}.  Q is smart.  Q is a socialist science sucker only because socialist science fails to sufficiently discourage the publication of inaccurate information by failing to hold A sufficiently accountable for being wrong.  Q is a socialist science sucker only because Q has had no other option.

Q now has another option.  Before launching a one year, one million dollar research project motivated by a result produced by socialist science, Q can spend ten days and ten thousand dollars on a few questions to check it out.  Before investing ten million dollars on a new materials science innovation, venture capitalists can anonymously conduct due diligence with a convenience and at a depth hitherto unimaginable.  Relying on free information of questionable accuracy produced by socialist science can be very costly.  Q is smart; Q is accustomed to paying for things of value; and Q's stakes are high.

Over time, Q increasingly relies on information obtained from capitalist science, where A has skin in the game.  When A has truly valuable information, he finds himself more interested in selling it than publishing it.  For less robust results, A is more inclined to publish, avoiding the harsh penalty imposed by capitalist science for providing information that turns out to be wrong.  Over time, useful, robust information slowly migrates to capitalist science, where it is appropriately valued.  Socialist science continues to publish the results A is willing to give away for free.

The forces in capitalist science's favor are so strong that socialist science should be allowed to fade into irrelevance in its natural course~\footnote{Although the information market unleashed by capitalist science could create millions of new science-related jobs, it would be irresponsible to reduce funding to socialist science until that promise has been realized.}.  Socialist science has been a remarkable institution.  The knowledge it has provided has been extraordinary in power and in scope.  Its use of an incentive structure spurned by the most developed of today's modern economies only makes its achievements all the more impressive.

We fully hope and expect socialist science will linger among us for many years.  At some deep level, the purpose and goals of socialist science are fundamentally different from those of capitalist science.  Capitalist science is designed to facilitate useful, accurate information transfer, and to provide A with a strong incentive to learn information some Q will find valuable.  Socialist science, in contrast, is not.

\bibliography{crisis_solution}

\end{document}